\documentclass[aps,prl,floatfix,twocolumn,amsfonts,amssymb,amsmath]{revtex4-2} 
\usepackage{float} 
\usepackage[colorlinks=true,linkcolor=black,citecolor=black,filecolor=black,urlcolor=black,breaklinks=true]{hyperref}
\usepackage{mathtools}
\usepackage[capitalize]{cleveref}
\usepackage{physics}
\usepackage{tikz}
\usepackage{multibib} 
\newcommand{\m}{\mathrm}

\begin{document}

\title{Measurement-Induced Quantum Synchronization and Multiplexing}
\author{Finn Schmolke}
\author{Eric Lutz}
\affiliation{Institute for Theoretical Physics I, University of Stuttgart, D-70550 Stuttgart, Germany}

\begin{abstract}
    Measurements are able to fundamentally affect quantum dynamics. 
    We here show that a continuously measured quantum many-body system can undergo a spontaneous transition from asynchronous stochastic dynamics to noise-free stable synchronization at the level of single trajectories.
    We formulate general criteria for this quantum phenomenon to occur, and demonstrate that the number of synchronized realizations can be controlled from none to all. We additionally find that ergodicity is typically broken, since time and ensemble averages may exhibit radically different synchronization behavior. 
    We further introduce a quantum type of multiplexing that involves individual trajectories with distinct synchronization frequencies. Measurement-induced synchronization appears as a genuine nonclassical form of synchrony that exploits quantum superpositions.
   \end{abstract}
   
   \maketitle 
   
   Synchronization is a universal concept in science and technology. Synchronous motion typically arises in coupled nonlinear oscillators when they collectively adjust their individual frequencies \cite{ble88,boc02,pik03,ace05,mos02,ani07,bal09,boc18}. Such phase-locking processes are omnipresent in physical, chemical, biological and engineering systems. Synchronization occurs in classical \cite{ble88,boc02,pik03,ace05,mos02,ani07,bal09,boc18} as well as in quantum \cite{goy06,zhi08,hei11,lud13,mar13,lee13,wal14,lor16,lor17,rou18,rou18a,son18,cab19,las20,sch22} systems. In both cases, three general synchronization mechanisms are commonly distinguished: Synchronized behavior may (i) spontaneously appear in interacting systems owing to their mutual coupling, (ii) be forced by an external periodic drive, or (iii) be induced by noise \cite{ble88,boc02,pik03,ace05,mos02,ani07,bal09,boc18,goy06,zhi08,hei11,lud13,mar13,lee13,wal14,lor16,lor17,rou18,rou18a,son18,cab19,las20,sch22}. However, classical and quantum theories are fundamentally different. An essential question is therefore to identify purely nonclassical forms of synchrony, determine their distinct quantum properties, and explore \mbox{their potential applications}.
   
   An intrinsic feature of quantum mechanics is measurement backaction that randomly perturbs the state of a measured system \cite{wis09,bar09,jac14}. As a consequence, observing a quantum object may dramatically affect its dynamics in a nonclassical manner. A case in point is the quantum Zeno effect, where frequent measurements slow down time evolution \cite{sud77}. While detection backaction is often detrimental, limiting measurement accuracy and causing decoherence \cite{wis09,bar09,jac14}, it has recently been realized that it may also be used to control complex many-body systems. Quantum measurements may indeed trigger phase transitions, such as entanglement phase transitions \cite{choi20,szy20,alb21,gop21,jia21,vov22,sie22,zab22,mul22,blo22,min22} and topological phase transitions \cite{geb20,sni21,wan22}, when the measurement rate, or strength, is varied. These measurement-induced phase transitions originate from the nontrivial interplay between unitary dynamics and the monitoring action of a detector, underscoring the potential constructive role of quantum measurements \cite{choi20,szy20,alb21,gop21,jia21,vov22,sie22,zab22,mul22,blo22,min22,geb20,sni21,wan22}.
   
   We here demonstrate measurement-induced synchronization in a continuously monitored quantum system. We concretely show that an otherwise closed many-body system may undergo a spontaneous transition from stochastic asynchronous dynamics to noise-free stable (anti)synchronization at the level of individual trajectories, when subjected to standard homodyne detection \cite{wis09,bar09,jac14}. 
   We formulate general criteria for measurement-induced synchronization in generic quantum systems based on the existence of decoherence-free subspaces \cite{dua97,zan97,lid98,lid03,blu08}. Decoherence-free subspaces are special parts of a system's Hilbert space that play an important role in quantum information science, since information encoded in them is protected from the environment \cite{dua97,zan97,lid98,lid03,blu08}. Whereas (anti)synchronization appears along all trajectories in classical systems, we show that the number of synchronized quantum trajectories is controlled by the overlap between the initial state and the decoherence-free subspaces. We reveal that synchronization may appear at the trajectory level, while being absent at the ensemble level---and vice versa. In general, knowledge of the ensemble average is hence not sufficient to provide information about the synchronized behavior of single realizations. We characterize this breaking of ergodicity by evaluating the fidelity between ensemble and time-averaged states \cite{joz94}. 
   We further introduce a quantum form of multiplexing \cite{zie09}, where individual synchronized trajectories at multiple frequencies are possible by engineering coherent superpositions of decoherence-free subspaces with distinct frequencies. Linear superposition, an essential resource of quantum theory \cite{nie00}, thus appears as a useful feature for quantum synchronization. Finally, we illustrate our results by analyzing measurement-induced synchronization in a quantum spin chain \cite{dut15}.
   
    \begin{figure*}[th]
    \centering
    \begin{tikzpicture}
    \node (a) [label={[label distance=-.34 cm]136: \textbf{(a)}}] at (0,0) {\includegraphics{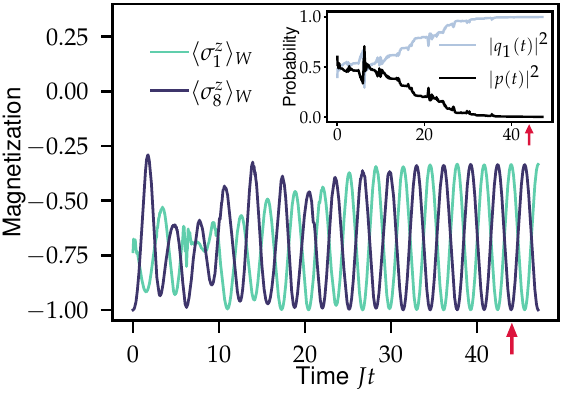}};	
    \node (b) [label={[label distance=-.34 cm]136: \textbf{(b)}}] at (5.6,0) {\includegraphics{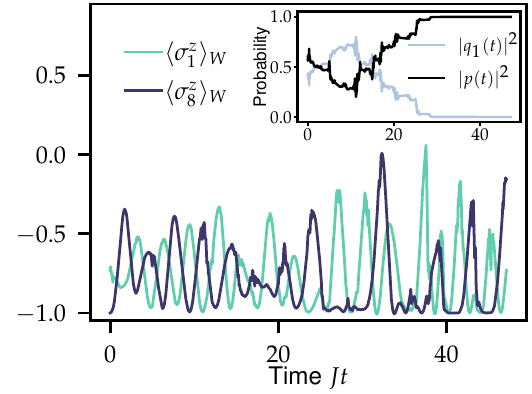}}; 
    \node (c) [label={[label distance=-.34 cm]136: \textbf{(c)}}] at (11.05,0) {\includegraphics{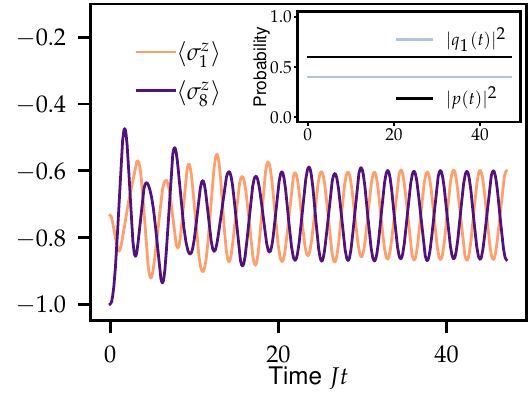}}; 
    \node (d) [label={[label distance=-.34 cm]136: \textbf{(d)}}] at (0,-4.1) {\includegraphics{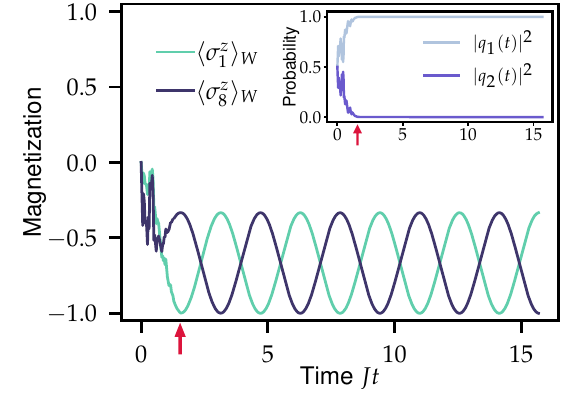}}; 
    \node (e) [label={[label distance=-.34 cm]136: \textbf{(e)}}] at (5.6,-4.1) {\includegraphics{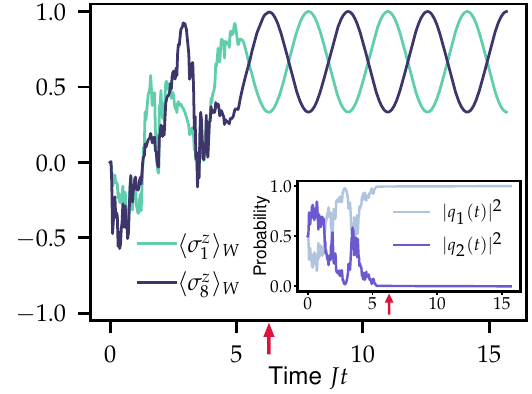}};
    \node (f) [label={[label distance=-.34 cm]136: \textbf{(f)}}] at (11.05,-4.1) {\includegraphics{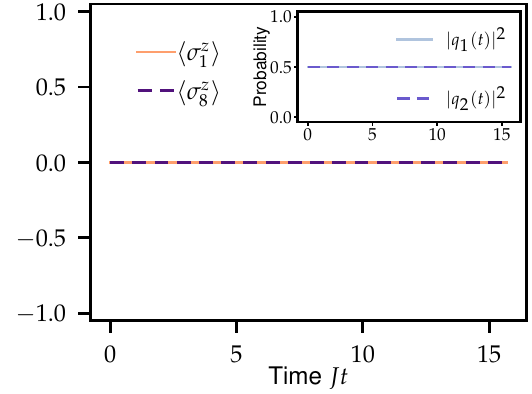}};
    \draw (a) node[yshift=2.3cm] {\textsf{Trajectory level}}; 
    \draw (b) node[yshift=2.3cm] {\textsf{Trajectory level}};
    \draw (c) node[yshift=2.3cm] {\textsf{Ensemble level}};
    \end{tikzpicture}
    \caption{Stable measurement-induced antisychronization along quantum trajectories. {Evolution of the end magnetizations, $\langle\sigma^z_1 \rangle_W$ and $\langle\sigma^z_8 \rangle_W$, of a $XY$ spin chain, \cref{chain}, with $N=8$ spins subjected to homodyne measurement of the third $z$-polarization with measurement operator $L = \sqrt{\Gamma} \sigma^z_3$. 
    (a) Spontaneous transition (red arrow) from noisy asynchronous to noiseless antisynchronization, when the system gets trapped in the first decoherence-free subspace with probability $|q_1|^2=1$}. The system is initially prepared in a mixture of the first decoherence-free subspace and the orthogonal complement, $|q_1(0)|^2=2/5$ and $|p(0)|^2=3/5$. 
    (b) No synchronization occurs, when the system gets trapped in the orthogonal complement with probability $|p|^2=1$. 
    (c) Antisynchronization also appears at the ensemble level for this configuration. 
    (d)-(e) When the state is initially prepared in a superposition of the two decoherence-free subspaces, $|q_1(0)|^2=1/2$ and $|q_2(0)|^2=1/2$, stable antisychronization occurs at the trajectory level, when the system gets spontaneously trapped in either one of the decoherence-free subspaces. 
    (f) However, in this case, there is no synchronization at the ensemble level. 
    The reduced measurement strength is $\Gamma/J = 0.7/\pi$.}
    \label{fig:homodyne}
   \end{figure*}
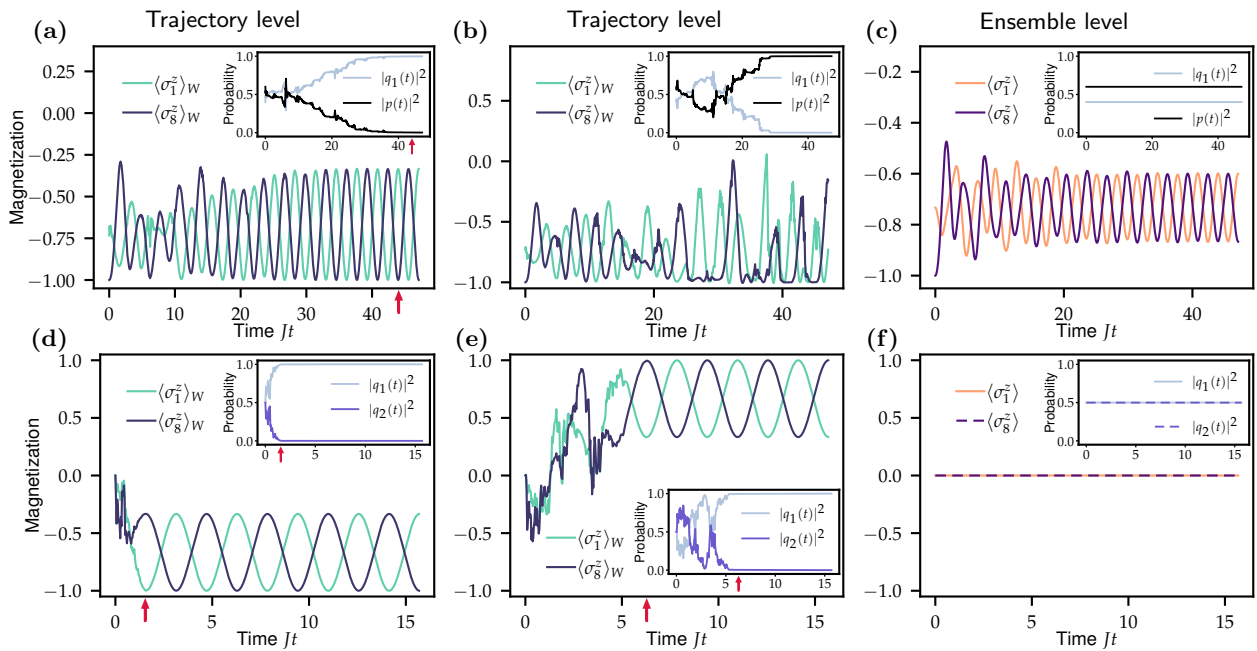
   
   \textit{Measurement-induced synchronization.} We consider a closed quantum system with Hamiltonian $H$ that is continuously monitored using a standard homodyne detection scheme \cite{wis09,bar09,jac14}. Its stochastic density operator $\rho_W$ evolves according to the It\^o stochastic master equation 
   \begin{align}
     \dd{\rho_W} =& -i[H,\rho_W] \dd{t} + \left(L\rho_W L^\dagger - \frac{1}{2}\left\{L^\dagger L,\rho_W\right\}\right) \dd{t} \notag \\
     &+ \left(L\rho_W+\rho_W L^\dagger - \langle L+L^\dagger\rangle \rho_W\right)\dd{W(t)},
     \label{eq:homodyne_sme}
   \end{align}
   where $L$ denotes the measurement operator and $\dd{W}$ is a Wiener noise increment satisfying $\dd{W(t)}^2=\dd{t}$ \cite{wis09,bar09,jac14}. We focus on a single measurement operator, but the analysis may easily be extended to an arbitrary number of them.
   The density operator $\rho_W$ describes a particular realization of the quantum process.
   Taking the ensemble average over all trajectories, \cref{eq:homodyne_sme} reduces to the usual Lindblad master equation,
   $\dot{\rho} = -i[H,\rho] + \mathcal{D}[L]\rho$, where $\rho= \mathbb{E}[\rho_W]$ is the averaged density operator, and 
    {$\mathcal{D}[L]\bullet = L\bullet L^\dagger - \left\{L^\dagger L,\bullet\right\}/2$} denotes the dissipator \cite{wis09,bar09,jac14}. We note that quantum synchronization has been mostly investigated at the ensemble level so far \cite{goy06,zhi08,hei11,lud13,mar13,lee13,wal14,lor16,lor17,rou18,rou18a,son18,cab19,las20,sch22}. The fluctuating statistics of synchronization have been examined for coupled mechanical oscillators (including nonlinear Van der Pol oscillators) by unraveling their dynamics in Refs.~\cite{wei16,es20}; in both cases, the evolution  is ergodic (since there is a unique steady state), and the corresponding trajectories remain stochastic.

   We begin by seeking conditions for stable measurement-induced (anti)synchronization, corresponding to local system observables that oscillate at the same frequency (and with stable amplitude), to occur along an individual quantum trajectory (we will see below that different trajectories may possibly exhibit different synchronization frequencies and amplitudes). This dynamical synchronization criterion has been widely used at the ensemble level \cite{gio12,gio13,kar19,gio19,kar20,tyn20,buc22}. 
   However, the averaged dynamics does not allow to draw conclusions  about individual realizations, which is why one needs to go beyond  existing conditions for quantum synchronization \cite{buc22} that are no longer applicable. 
   In particular, the occurrence of synchronization in the mean does generally not imply synchronization along single trajectories, and vice versa. 
   Typically, the temporal behavior of a continuously monitored system remains stochastic throughout the whole evolution.
   Thus, to ensure stable synchronization, we require the onset of unperturbed oscillations with constant amplitude. 
   A sufficient condition is the existence of a decoherence-free subspace (DFS), such that $L \ket{q_k} = c_k \ket{q_k}$, where $\ket{q_k}$ are eigenstates of the Hamiltonian and $c_k$ are complex numbers \cite{dua97,zan97,lid98,lid03,blu08} (different constants $c_k$ generally correspond to distinct decoherence-free subspaces). This follows from the observation that a decoherence-free subspace remains decoherence-free along a trajectory \eqref{eq:homodyne_sme}, since $\dd{\rho^\m{DFS}_W} = -i[H,\rho^\m{DFS}_W] \dd{t}$. 
   Contrary to (averaged) Lindblad dynamics, where the state space needs to be considered as a whole, here, each (decoherence-free) subspace must be treated independently to be able to properly account for synchronization along an individual trajectory.
   A sufficient condition for stable (anti)synchronization is thus that the decoherence-free subspace only contains a single eigenmode, in which case (anti)synchronization appears at the \mbox{frequency of that eigenmode \cite{com}}.
   
   We next examine the dynamics of the  synchronization process by first assuming the presence of a single decoherence-free subspace. To that end, we show that, when the measurement operator is Hermitian, the probability that individual realizations get spontaneously trapped in the decoherence-free subspace is equal to the initial support on that subspace. We start by writing the solution of the stochastic master equation \eqref{eq:homodyne_sme}  as 
   $
       \rho_W(t) = \sum_m u_m \dyad{\Psi_m(t)}
   $,
   where $ u_m $ is the probability to prepare the pure state $\ket{\Psi_m(0)}$.
   We further partition the system Hilbert space  into the decoherence-free subspace (with basis states $\{\ket{q_k}\}$) and its orthogonal complement (with basis states $\{\ket{p_l}\}$). Any pure   state $\ket{\Psi(t)}$ may then be expanded as
   $ \ket{\Psi(t)} = \sum_{k} q_k(t) \ket{q_k} + \sum_{l} p_l(t) \ket{p_l}$, with
    $q_k(t) = \braket{q_k}{\Psi(t)}$ and $p_l(t) = \braket{p_l}{\Psi(t)}$ (we omit the state index $m$ for convenience).
   The probabilities to find the system  at {time $t$ in the decoherence-free subspace or its complement are thus respectively    $ |q(t)|^2 = \sum_k |q_k(t)|^2$
   and $ |p(t)|^2 = \sum_l |p_l(t)|^2 = 1- |q(t)|^2$, since  total probability is conserved. By additionally assuming  that $L = L^\dagger$, we obtain  using Eq.~\eqref{eq:homodyne_sme} the differential for the probability $|q(t)|^2$ (Supplemental Material \cite{sup})
   \begin{align}
       \dd{(|q|^2)} = 2|q|^2 \bigg(c(t)\left(1-|q|^2\right) 
        - \sum_{m,n} p^\ast_mp_n L_{mn}\bigg)\dd{W},
       \label{eq:dq}
   \end{align}
   where we have defined $\sum_k |q_k(t)|^2 c_k \equiv |q(t)|^2c(t)$ and $L_{mn} = \bra{p_m}L\ket{p_n}$. \Cref{eq:dq} describes the temporal evolution of the overlap with the decoherence-free subspace for individual realizations. It has the form of a free Brownian motion with state-dependent diffusion \cite{kam92}. 
   The corresponding Fokker-Planck equation for the probability density $P(|q|^2,t)$ is accordingly \cite{kam92}
   \begin{equation}
    \!\! \! \frac{\partial P}{\partial t} = 2\pdv[2]{(|q|^2)} \Bigg[|q|^4\bigg(\!c(t)\left(1-|q|^2\right)     - \sum_{m,n} p^\ast_mp_n L_{mn}\bigg)^2 \!P\Bigg]\!\!
       \label{eq:fpe}
   \end{equation}
   with  steady-state solution  (Supplemental Material \cite{sup})
   \begin{align}
       {P^\m{s}}(|q|^2) = \left(1-|q(0)|^2\right)\delta(|q|^2) + |q(0)|^2\delta(|q|^2-1). 
       \label{eq:stat-pdf}
   \end{align}
   A single stochastic trajectory will therefore asymptotically {select} the decoherence-free subspace, $|q|^2=1$, (and become unitary)  with probability $|q(0)|^2$, or the orthogonal complement, $|q|^2=0$,  (and remain stochastic) with probability $1-|q(0)|^2$. When a  trajectory gets trapped indefinitely in  a decoherence-free subspace, the quantum system   undergoes a spontaneous transition from random to noiseless evolution, which may support (anti)synchronization. These results can be easily extended to arbitrary mixed states and multiple decoherence-free subspaces (Supplemental Material \cite{sup}). The above scenario bears similarities with dissipative freezing  which has recently been investigated for  quantum jump processes \cite{san19,hal22,tin23}. However, dissipative freezing entails freezing into arbitrary symmetry sectors where the evolution is generally stochastic, and stable synchronization is therefore absent.

   \begin{figure}[t]
    \centering
    \begin{tikzpicture}
     \node (a) [label={[label distance=-.45 cm]130: \textbf{a)}}] at (0,0) {\includegraphics{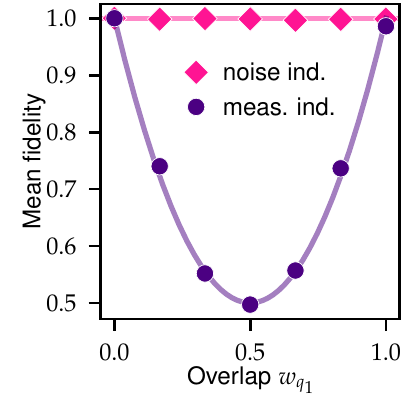}};	\node (a) [label={[label distance=-.45 cm]130: \textbf{b)}}] at (4.3,0) {\includegraphics{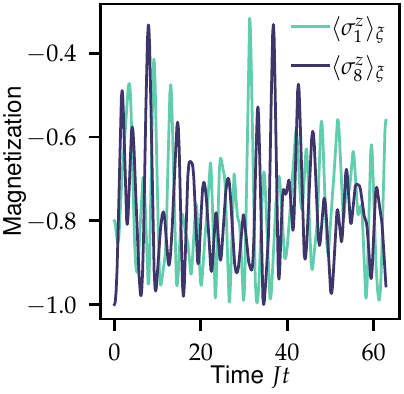}};
     \end{tikzpicture}
    \caption{Ergodicity breaking. 
     (a) Mean fidelity between time and ensemble-averaged states, \cref{eq:ipr}, as a function of the overlap, $w_{q_1} = |q_1(0)|^2$, with the first decoherence-free subspace, for quantum (measurement) noise (purple line). 
     Dynamics is nonergodic unless the chain starts in one of the subspaces (dots show simulations with 100 trajectories). Dynamics is always ergodic for classical noise (pink). 
     (b) Corresponding trajectories are not synchronized (shown for $w_{q_1}=0.3$).}
    \label{fig:ergodicity}
   \end{figure} 
   
   \begin{figure*}[ht]
    \centering
    \begin{tikzpicture}
     \node (a) [label={[label distance=-.3 cm]136: \textbf{(a)}}] at (0,0) {\includegraphics{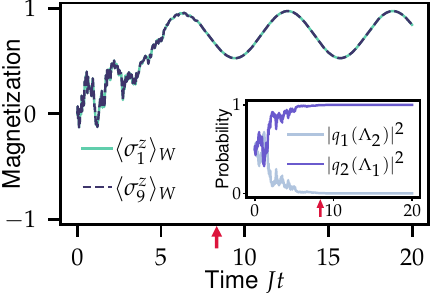}};	
     \node (a) [label={[label distance=-.3 cm]136: \textbf{(b)}}] at (4.4,0) {\includegraphics{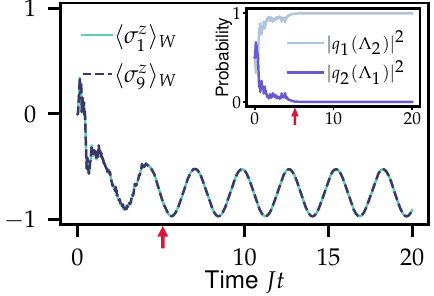}}; 
     \node (a) [label={[label distance=-.3 cm]136: \textbf{(c)}}] at (8.8,0) {\includegraphics{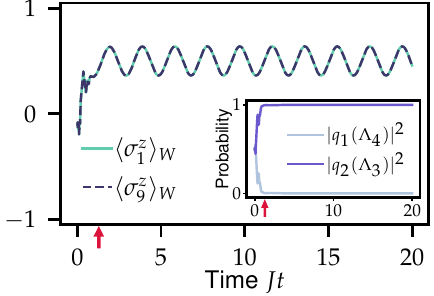}}; 
     \node (a) [label={[label distance=-.3 cm]136: \textbf{(d)}}] at (13.2,0) {\includegraphics{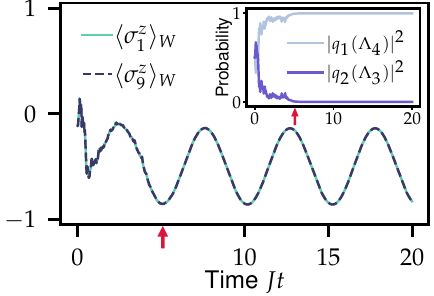}}; 
     \end{tikzpicture}
    \caption{Quantum multiplexing and stable measurement-induced synchronization. Evolution of the end spin magnetizations, $\langle\sigma^z_1 \rangle_W$ and $\langle\sigma^z_9 \rangle_W$, of a $XY$ spin chain, \cref{chain}, with $N=9$ spins, subjected to homodyne measurement of the fifth $z$-polarization with measurement operator $L = \sqrt{\Gamma} \sigma^z_5$. The chain is prepared in a linear superposition of two decoherence-free subspaces that support the same four eigenfrequencies, $\Lambda_1 = J$, $\Lambda_2 = \sqrt{5}J$, $\Lambda_3 = [\sqrt{5}+1]J$ and $\Lambda_4 = [\sqrt{5}-1]J$. 
    (a)-(b) When the system is initialized in $\ket{\Psi(0)} = (\ket{q_1(\Lambda_2)}+\ket{q_2(\Lambda_1)})/\sqrt{2}$, stable synchronization  occurs with eigenfrequencies $\Lambda_{2(1)}$ in the subspace with $c_{1(2)}$.
    (c)-(d) When the system is initialized in $\ket{\Psi(0)} = (\ket{q_1(\Lambda_4)}+\ket{q_2(\Lambda_3)})/\sqrt{2}$, stable synchronization occurs  with eigenfrequencies $\Lambda_{4(3)}$ in the subspace with $c_{1(2)}$. The reduced measurement strength is $\Gamma/J = 0.1$.}
    \label{fig:multiplexing}
   \end{figure*}
   
   \textit{Example of a quantum spin chain.}
   The above discussion is valid for generic  quantum many-body systems. 
   We now apply our findings to a $XY$-chain of $N$ spins in a transverse field \cite{dut15}. 
   The corresponding Hamiltonian is 
   \begin{align}
   \label{chain}
    H = \frac{J}{2}\sum_{j = 1}^{N-1} (\sigma^x_j\sigma^x_{j+1}+\sigma^y_j\sigma^y_{j+1}) + h \sum_{j = 1}^N \sigma^z_j, 
   \end{align}
   where $\sigma_j^{x,y,z}$ are the usual Pauli operators, $J$ is the interaction constant, and $h = 1$ is the field strength. 
   The quantum $XY$-chain is an important system in statistical physics and quantum information science \cite{dut15}. 
   For concreteness, we consider a chain of $N=8$ spins and perform a homodyne measurement of the polarization of the third spin by setting $L = \sqrt{\Gamma}\sigma^z_3$, where $\Gamma$ is the measurement strength. This model admits two two-dimensional decoherence-free subspaces with $c_1 = -1$ and $c_2 = 1$ (Supplemental Material \cite{sup}). 
   Both subspaces support antisynchronization, at the same frequency, of the average polarizations, $\langle \sigma^z_{1}\rangle_W$ and $\langle \sigma^z_{8}\rangle_W$, of the two edge spins, with $\langle \sigma^z_{j}\rangle_W =\Tr[\sigma^z_{j} \rho_W]$ (the dynamics of the remaining magnetizations are presented in the Supplemental Material \cite{sup}). 
   To illustrate the nonintuitive behavior of the system and highlight its quantum properties, we compare {two scenarios}: (i) in the first case, the system is initially in a statistical mixture of the first decoherence-free subspace and the orthogonal complement, $|q_1(0)|^2 = 2/5$, $|q_2(0)|^2 = 0$, $|p(0)|^2 = 3/5$ (Figs.~\ref{fig:homodyne}abc), whereas (ii) in the second case, the initial state is chosen {such} that the chain is in a linear superposition of the first and second decoherence-free subspace, $|q_1(0)|^2 = |q_2(0)|^2 = 1/2$, $|p(0)|^2 = 0$, (Figs.~\ref{fig:homodyne}def). 
   Figure~\ref{fig:homodyne}a demonstrates measurement-induced stable quantum antisynchronization, where the first and the last spin oscillate with identical frequencies. As predicted by Eqs.~\eqref{eq:fpe}-\eqref{eq:stat-pdf}, two fifths of the trajectories undergo a spontaneous transition (inset) from random asynchronous to noiseless antisynchronized evolution, as they get trapped in the first decoherence-free subspace. The remaining trajectories, by contrast, end up in the orthogonal complement, stay noisy and do not synchronize (Fig.~\ref{fig:homodyne}b). 
   The two end spins also exhibit antisynchronization at the ensemble level, with average polarizations $\langle \sigma^z_{j}\rangle =\Tr[\sigma^z_{j} \rho]$ (\cref{fig:homodyne}c). On the other hand, in case (ii), half of the trajectories get localized in each decoherence-free subspace, implying that antisynchronization occurs along all realizations (\cref{fig:homodyne}de). However, here antisynchronization is completely absent at the ensemble level (\cref{fig:homodyne}f), indicating that individual realizations can strongly deviate from the mean. 
   This genuinely nonclassical phenomenon is a consequence of quantum superposition. 
   It may be experimentally observed in superconducting circuits that have been used to realize spin chains \cite{guo21,mi22} and monitor individual trajectories of continuously measured qubits \cite{mur13,web14}.

   \textit{Ergodicity breaking and classical noise.}
   The  identified difference between trajectory and ensemble properties  is related to ergodicity breaking \cite{pal82}. 
   We quantify nonergodic behavior with the mean fidelity, $\mathbb{E}[F\left(\overline{\rho}_W,\rho^\m{s}\right)] = \mathbb{E}\Big[\Tr[\sqrt{\sqrt{\overline{\rho}_W} \rho^\m{s}\sqrt{\overline{\rho}_W}}]^2\Big]$, between the time-averaged state, $\overline{\rho}_W = \lim_{T \to \infty} \int_{0}^{T} \dd{t}\, \rho_W/T$, and the ensemble-averaged state $\rho^\m{s}$ \cite{joz94}, for a {given} initial condition.
   When one decoherence-free subspace exists, and the operator $L$ is Hermitian, the system Hilbert space may be decomposed into two mutually orthogonal subspaces, that can be associated to orthogonal blocks of the density matrix \cite{bau12}. When more decoherence-free subspaces are present, additional orthogonal subspaces may be identified.
   We assume that every block $j$ has a unique steady state $\rho^\m{s}_j$. These asymptotic states are orthogonal \cite{bau08,bau12}.
   The ensemble-averaged density matrix can then be written as 
   $
    \rho ^\m{s} = \sum_j w_j \rho^\m{s}_j,
   $ 
   where $w_j$ is the probability that the initial state is prepared in block $j$.
   Since each block contains exactly one stationary state $\rho^\m{s}_j$, every trajectory localizes into one of them with probability $w_j$, where the evolution is ergodic \cite{kum04}. As a result,
   $
    \overline{\rho}_{W,j} = {\rho^\m{s}_j}
   $, and we obtain \cite{com1}
   \begin{equation}
   \!\! \mathbb{E}[F\left(\overline{\rho}_{W},\rho^\m{s}\right)]
    = \sum_k w_k F\bigg(\overline{\rho}_{W,k},\sum_j w_j \rho^\m{s}_j\bigg)
    = \sum_k w_k^2.
    \label{eq:ipr}
   \end{equation}
   The mean fidelity is thus given by the inverse participation ratio, a prominent measure of localization \cite{kra93,eve08}, of the initial state over the subspaces. Equation \eqref{eq:ipr} is lower bounded by $1/N$, where $N$ is the number of blocks (Supplementary information). The dynamics is accordingly ergodic only when the system starts in one of the subspaces.
   \Cref{fig:ergodicity}a displays the mean fidelity \eqref{eq:ipr} as a function of the overlap with the first decoherence-free subspace, $|q_1(0)|^2$, for the example of the quantum spin chain \eqref{chain}. Taking the initial state, $\rho(0) = w_{q_1} \dyad{q_1} + w_p \dyad{p}$, with $w_{q_1}= |q_1(0)|^2$ and $w_p = |p(0)|^2$, the mean fidelity is simply given by $\mathbb{E}[F\left(\overline{\rho}_W,\rho^\m{s}\right)] = w_{q_1}^2 + (1-w_{q_1})^2$ (purple line), in perfect agreement with numerical simulations of the quantum trajectories (purple dots).
   
   It is instructive to further compare synchronization induced by quantum (measurement) noise and by classical noise \cite{sch22}. Setting $L = -i\sqrt{\Gamma}\sigma^z_3$, the randomness no longer depends on the state of the system, leading to stochastic unitary dynamics, $\dot{\rho}_\xi = -i [H + \sqrt{\Gamma}\xi(t)\sigma^z_3,\rho_\xi]$ (in Stratonovich convention), with classical white noise with zero mean and unit variance $\langle\xi(t)\xi(t^\prime)\rangle = \delta(t-t^\prime)$ \cite{bar09}. 
   Figure \ref{fig:ergodicity}a shows that for classical noise, the evolution is always ergodic (pink line and diamonds) for any initial overlap $|q_1(0)|^2$ with the first decoherence-free subspace. 
   In this situation, there is no (anti)synchronization along single trajectories (\cref{fig:ergodicity}b), although for finite $|q_1(0)|^2$, synchronous behavior appears at the ensemble level \cite{sch22}. 
   The effects of classical and quantum noises on the system hence differ significantly.
   
   \textit{Application to quantum multiplexing}.
   Multiplexing is a standard technique in classical communication by which signals with different frequencies are simultaneously transmitted through the same medium \cite{zie09}. 
   Measurement-induced synchronization allows for a quantum form of multiplexing by preparing the system in a superposition of multiple decoherence-free subspaces with distinct frequencies. According to \Cref{eq:stat-pdf}, the initial overlap with each subspace then controls the amount of trajectories that are (anti)synchronized at the respective frequencies. 
   \Cref{fig:multiplexing} presents an example of quantum multiplexing using the quantum $XY$ chain, \cref{chain},
   with $N = 9$ spins and the polarization of the fifth spin $L = \sqrt{\Gamma}\sigma^z_5$, being measured. 
   Contrary to the previous example, this system supports stable  synchronized trajectories at various frequencies. 
   It indeed possesses two eight-dimensional decoherence-free subspaces with the same eigenmodes, and eigenfrequencies given by, $\Lambda_1 = J$, $\Lambda_2 = \sqrt{5}J$, $\Lambda_3 = [\sqrt{5}+1]J$, $\Lambda_4 = [\sqrt{5}-1]J$. 
   To implement multiplexing, we prepare a superposition of the two subspaces, $\ket{\Psi(0)} = (\ket{q_1(\Lambda_j)}+\ket{q_2(\Lambda_k)})/\sqrt{2}$, where $\ket{q_i(\Lambda_j)}$ indicates a state with eigenfrequency $\Lambda_j$ in the subspace belonging to $c_i$. 
   Depending on the choice of the eigenmodes, one half of the trajectories can be  synchronized at frequency $\Lambda_j$ and the other half at frequency $\Lambda_k$ (\cref{fig:multiplexing}abcd). 
   More frequencies could be superposed in configurations with more decoherence-free subspaces.

   \textit{Conclusions.} Classical synchronization plays an important role for classical communication systems \cite{arg05,cho17}. 
   The study of nonclassical types of synchrony and the exploration of their potential applications for quantum communication purposes are hence of great interest. 
   We have here analyzed quantum synchronization induced by continuous (homodyne) measurements. 
   In particular, we have shown that a many-body system may undergo a spontaneous transition from random asynchronous dynamics to stable noise-free (anti)synchronization at the level of individual trajectories. 
   The number of (nonergodic) synchronized realizations is given by the initial overlap with a decoherence-free subspace, and can thus be controlled by preparing a linear superposition of states living in different subspaces. 
   A quantum form of frequency multiplexing is consequently possible, when these decoherence-free subspaces are associated with different eigenfrequencies. These results highlight the significance of coherent superpositions for quantum synchronization.

   We acknowledge support from the Vector Foundation.

\clearpage
\widetext
\begin{center}
\textbf{\large Supplemental Material: Measurement-Induced Quantum Synchronization and Multiplexing}
\end{center}
\setcounter{equation}{0}
\setcounter{figure}{0}
\setcounter{table}{0} 
\setcounter{page}{1}
\makeatletter
\renewcommand{\theequation}{S\arabic{equation}}
\renewcommand{\thefigure}{S\arabic{figure}}
\renewcommand{\bibnumfmt}[1]{[S#1]}
\renewcommand{\citenumfont}[1]{S#1}


\renewcommand{\figurename}{Supplementary Figure}
\renewcommand{\theequation}{S\arabic{equation}}
\renewcommand{\thefigure}{S\arabic{figure}}
\renewcommand{\bibnumfmt}[1]{[S#1]}
\renewcommand{\citenumfont}[1]{S#1}
\renewcommand{\thesection}{\Roman{section}} 

This Supplemental Material provides additional information about (I) the derivation of the evolution equation of the probability to find the system in a decoherence-free subspace, (II) the determination of the corresponding stationary probability distribution (III) its extension to mixed states, as well as (IV) the identification of the decoherence-free subspaces of the quantum $XY$ spin chain, (V) the behavior of the synchronization time, and (VI) the evaluation of the mean fidelity that characterizes the breaking of ergodicity.

\section{Differential for the probability of being in a subspace of the Hilbert space}
In this section, we derive the equation of motion for the probability $|q(t)|^2$ that the system is found in the decoherence-free subspace at time $t$, as given in Eq.~(2) of the main text.
A pure state of a quantum system subject to homodyne detection  evolves according to the stochastic Schr\"odinger equation \cite{wis09,bar09,jac14}
\begin{align}
  \dd{\ket{\Psi(t)}} 
  &= \left(-iH \dd{t} -  \left[\frac{L^\dagger L}{2} - \frac{X(t)L}{2} + \frac{X^2(t)}{8}\right]\dd{t} + \left[L-\frac{X(t)}{2}\right]\dd{W}(t)\right) \ket{\Psi(t)},
  \label{seq:sse}
\end{align}
where $X(t) = \Tr[(L+L^\dagger)\dyad{\Psi(t)}]$.
The Hilbert space may  be partitioned into the decoherence-free subspace (with basis states $\{\ket{q_k}\}$) and its orthogonal complement (with basis states $\{\ket{p_k}\}$). Any pure state $\ket{\Psi(t)}$ may then be expanded as
\begin{align}
\ket{\Psi(t)} = \sum_{k} q_k(t) \ket{q_k} + \sum_{l} p_l(t) \ket{p_l}, 
\label{seq:bipartition}
\end{align}
with
$q_k(t) = \braket{q_k}{\Psi(t)}$ and $p_l(t) = \braket{p_l}{\Psi(t)}$. 
For every pure state $\ket{\Psi(t)}$, we can compute the probability for the system to be found in the decoherence-free subspace at time $t$ as the overlap $|q(t)|^2 = \sum_k |\braket{q_k}{\Psi(t)}|^2$. The differential then follows by It\^o  calculus \cite{kam92} as
\begin{align}
  \dd{\big(|q(t)|^2\big)} 
  = \sum_k \bra{q_k} \big[\dyad{\dd{\Psi(t)}}{\Psi(t)} +  \dyad{\Psi(t)}{\dd{\Psi(t)}} + \dyad{\dd{\Psi(t)}}{\dd{\Psi(t)}}\big] \ket{q_k}.
\end{align}
This expression may be evaluated by inserting  \cref{seq:sse}. To this end, we need the assumption that the measurement operator $L$ is Hermitian, which guarantees that the decoherence-free subspace is orthogonal, i.e $\bra{q_k}L\ket{p_l} = \bra{p_k}L\ket{q_l} = 0$. As a result, the deterministic terms cancel and only the stochastic term remains. This leads to
\begin{align}
  \dd{\big(|q(t)|^2\big)} 
  &= \left(\sum_k q^\ast_k(t)\bra{q_k} \left[L-\frac{X(t)}{2}\right] \ket{\Psi(t)} 
  + q_k(t) \bra{\Psi(t)}\left[L^\dagger - \frac{X(t)}{2}\right] \ket{q_k}\right) \dd{W}\\
  &= 2 |q(t)|^2 \bigg(c(t)\left(1-|q(t)|^2\right) 
  - \sum_{m,n} p^\ast_m(t)p_n(t) L_{mn}\bigg)\dd{W}.
  \label{seq:dq}
\end{align}
To arrive at the second line, we have used the decomposition of $\ket{\Psi(t)}$ into the orthogonal subspaces and introduced the function $c(t)$ through the relation $\sum_k |q_k(t)|^2c_k \equiv |q(t)|^2c(t)$ where $L_{mn} = \bra{p_m}L\ket{p_n}$. 

If multiple decoherence-free subspaces exist, the bipartition in \cref{seq:bipartition} may still be performed where, in the first step, all decoherence-free subspaces are combined in the term $|q(t)|^2$.
When the system then enters this subspace, a further bipartition of the remaining space can be made, yielding 
\begin{align}
    \dd{(|q_1(t)|^2)} = 
    2 |q_1(t)|^2 \left(1-|q_1(t)|^2\right)[c_1(t)- c_2(t)]\dd{W},
    \label{seq:dq_12} 
\end{align}
with $|q_1(t)|^2 + |q_2(t)|^2 = 1$ and $\sum_k |q_{x,k}(t)|^2 c_{x,k} \equiv c_x(t) |q_x(t)|^2$, where $x = 1,2$.
The line of arguments following \cref{seq:dq} still applies and ensures that, eventually, one of the two subspaces is selected.
This procedure can be repeated until no further partition into distinct decoherence-free subspaces is possible.
Hence, in the presence of multiple decoherence-free subspaces, only one of them is ever selected, in which the system will then remain indefinitely.

\section{Steady-state distribution}
We next derive the steady-state probability distribution for $|q(t)|^2$, Eq.~(4) of the main text. This may be done by using the asymptotic limit of the 
the Fokker-Planck equation for $P(|q(t)|^2,t)$. However, it is more convenient to approach the problem from the trajectory description given in \cref{seq:dq}. We first evaluate the mean value,
   $ \mathbb{E}\left[\dd{\big(|q(t)|^2\big)}\right] = 0 $, using \cref{seq:dq}, from which we can conclude that     $\mathbb{E}\left[|q(t)|^2\right] = |q(0)|^2$,
for all $t$, where $\mathbb{E}[\bullet]$ denotes the expectation value over the ensemble of measurement realizations.
We subsequently note that the two fixed points of the dynamics, $|q(t)|^2 \in \{0,1\}$, can be immediately identified by inspection: 
For generic operators $H,L$, the first and the second term inside the parentheses on the rhs of \cref{seq:dq} are indeed independent and the two fixed points $|q(t)|^2 \in \{0,1\}$ are the only two stationary solutions of \cref{seq:dq}.
To  produce the mean value $|q(0)|^2$, the amount of trajectories that reach $|q(t)|^2 = 1$ has to be equal to $|q(0)|^2$.
The stationary probability distribution is consequently given by 
\begin{align}
  P^\m{s}(|q|^2) = \left(1-|q(0)|^2\right)\delta(|q|^2) + |q(0)|^2\delta(|q|^2-1). 
  \label{seq:stat-pdf}
\end{align}

\begin{figure}[t]
    \centering
    \includegraphics{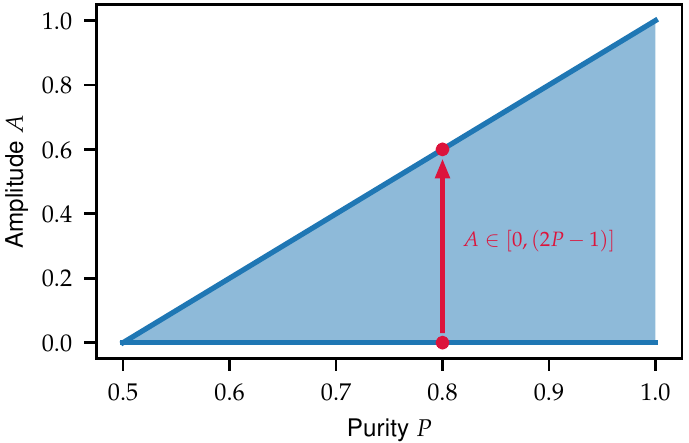}
    \caption{Amplitude of synchronized oscillations as a function of the purity of the initial state. The amplitude $A$ is upper bounded by $2P-1$, but its value is not directly related to that of the purity.}
    \label{fig:purity}
\end{figure}

\section{Extension to mixed states}
Our results  hold for arbitrary mixed states, since the overlap of the initial state with a decoherence-free subspace does not depend on its purity. To show this, we write the solution of the stochastic differential equation \eqref{seq:sse} for an initial state $\rho(0) = \sum_m u_m \dyad{\Psi_m(0)}$  as 
\begin{align}
  \rho(t) = \sum_m u_m \dyad{\Psi_m(t)},
\end{align}
where each of the pure states $\ket{\Psi_m(t)}$ evolves according to the stochastic Schr\"odinger equation (S1). 
The probabilities to find the system either in the decoherence-free subspace or its orthogonal complement are then 
\begin{align}
  |q(t)|^2 &= \Tr[\rho(t)P_q] = \sum_{m,k} u_m |\braket{q_k}{\Psi_m(t)}|^2,\\
  |p(t)|^2 &= \Tr[\rho(t)P_p] = \sum_{m,k} u_m |\braket{q_k}{\Psi_m(t)}|^2,
\end{align}
where $P_q$ and $P_p$ are projectors onto the decoherence-free subspace and its complement, respectively.
For mixed states, we must thus additionally account for the probability with which each of the pure states is prepared.
The remaining derivation then follows analogously to the one described in the main text.
We accordingly obtain for the stationary probabilities to end up in the decoherence-free subspace and its complement the respective expressions
\begin{align}
    P^\mathrm{s}(\rho \in \mathrm{DFS}) &= \Tr[\rho(0)P_q] = |q(0)|^2,\\
    P^\mathrm{s}(\rho \notin \mathrm{DFS}) &= \Tr[\rho(0)P_p] = |p(0)|^2,
\end{align}
This is again the initial support of the density matrix in each of the corresponding subspaces. It thus follows that the stationary probability distribution is Eq.~(4) of the main text
\begin{align}
    P^\mathrm{s}(|q|^2) = (1-|q(0)|^2)\delta(|q|^2) + |q(0)|^2\delta(|q|^2-1).
\end{align}

However, the amplitude of the synchronized oscillations depends on the purity. They will therefore disappear in the classical limit. Once trapped in a decoherence-free subspace, the evolution is effectively unitary, and, thus, whenever synchronization occurs, the quality of that synchronization is always perfect (in the sense that the oscillations are perfectly periodic and the amplitude is stable). To better understand the relationship between the amplitude of synchronized oscillations and the purity of the initial state, we focus on a two-dimensional decoherence-free subspace. We  consider a single qubit with Hamiltonian $H = \sigma^z$ and examine  the unitary evolution of the observable $\sigma^x$. This model is analogous to the effective evolution inside the decoherence-free subspace.

The initial state in the Bloch representation is
\begin{align}
    \rho(0) = \frac{1}{2}
    \begin{pmatrix}
        1+a_z & a_x-i a_y\\
        a_x+ia_y & 1-a_z
    \end{pmatrix},
\end{align}
with Bloch vector $\vb*{a} = (a_x,a_y,a_z)^\mathrm{T}$.
The purity and ($l_1$-norm of) coherence are respectively given by \cite{str17}
\begin{align}
    P &= \Tr[\rho^2(0)] = \frac{1}{2}(1+|\vb*{a}|^2),\\
    C_{l_1} &= 2\sqrt{a_x^2+a_y^2}.
\end{align}
The expectation value of $\sigma^x$ now evolves according to
\begin{align}
    \Tr[\sigma^x \rho(t)] = \cos(2t)a_x - \sin(2t)a_y,
\end{align}
with an oscillation amplitude $A = \sqrt{a_x^2+a_y^2}$, which is exactly half the coherence.
By combining the above expressions, we may further write the amplitude as a function of the purity
\begin{align}
\label{eq}
    A = \sqrt{(2P-1)^2- {a_z^2}}.
\end{align}
Equation \eqref{eq} shows that the amplitude is upper bounded by $2P-1$, as can be seen in \cref{fig:purity}. We note that a general tradeoff between purity and mixedness has been demonstrated in Ref.~\cite{sin15}.

\section{Decoherence-free subspaces of the $\boldsymbol{XY}$-chain}
We here show how to identify the decoherence-free subspaces of the $XY$-chain with Hamiltonian 
\begin{align}
  H = \frac{J}{2}\sum_{j = 1}^{N-1} (\sigma^x_j\sigma^x_{j+1}+\sigma^y_j\sigma^y_{j+1}) + h \sum_{j = 1}^N \sigma^z_j,
  \label{seq:chain} 
\end{align}
and Lindblad jump operator $L = \sqrt{\Gamma}\sigma^z_u$, where $\Gamma$ is the measurement strength. 
The corresponding Lindblad equation can be written in vectorized form in Liouville space \cite{gya20}, where the dissipator is then treated as a perturbation. Applying perturbation theory, one may obtain the decay rates for each eigenmode as discussed in Ref.~\cite{sch22}. Eigenmodes with zero decay rates are located in a decoherence-free subspace. With this approach, we are able to find the configurations that admit decoherence-free subspaces and obtain the corresponding states $\ket{q_k}$ and eigenvalues $\lambda_j$. 
These states may still belong to different subspaces.
To identify those, we  compute the constants $c_k = \bra{q_k}L\ket{q_k}$, since different constants $c_k$ belong to different subspaces. 
Using this procedure, for $N = 8$ spins and $L = \sqrt{\Gamma}\sigma^z_3$, we identify two decoherence-free subspaces with $c_1 = -1$ and $c_2 = 1$, both supporting the same eigenfrequency $\Lambda = 2J$. 

On the other hand, for $N = 9$ qubits with jump operator $L = \sqrt{\Gamma}\sigma^z_5$, the problem has a reflection symmetry with respect to the middle of the chain. Therefore, in the asymptotic regime $t \gg 1/\Gamma$. we have $\langle \sigma^z_j\rangle(t) = \langle \sigma^z_{N+1-j} \rangle(t)$ for $j \in [1,4]$.
Because of the high degree of symmetry, the decoherence-free subspace is much larger. However, since the only eigenvalues of $\sigma^z_u$ are $\pm 1$, there are still only two distinct decoherence-free subspace with $c_1 = 1$ and $c_2 = -1$, both eight-dimensional. We concretely find the four eigenfrequencies, $\Lambda_1 = J$, $\Lambda_2 = J\sqrt{5}$, $\Lambda_3 = J\left[\sqrt{5}+1\right]$, $\Lambda_4 = J\left[\sqrt{5}-1\right]$, supported in both subspaces.
\begin{figure}[ht]
\centering
 \begin{tikzpicture}
  \node (a) [label={[label distance=-0.6 cm]162: \textbf{(a)}}] at (0,0) {\includegraphics{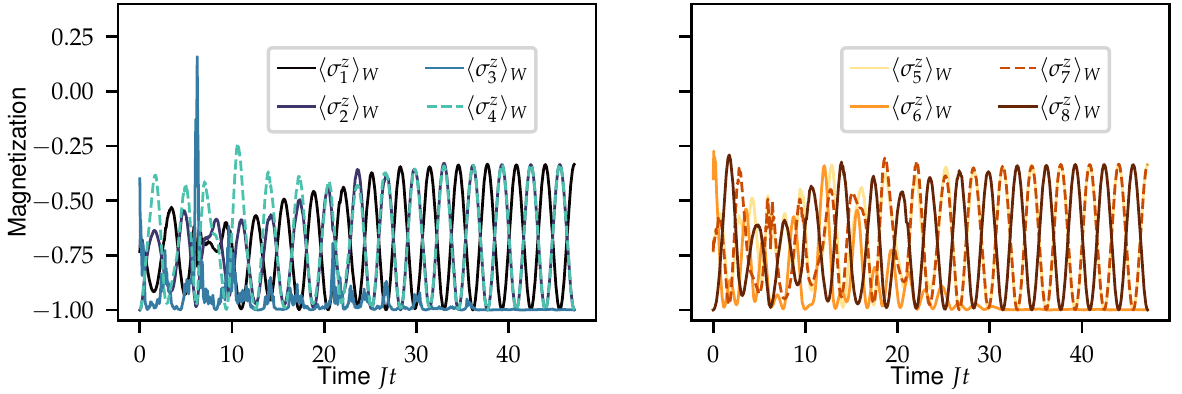}};	
  \node (b) [label={[label distance=-0.6 cm]162: \textbf{(b)}}] at (0,-4.4) {\includegraphics{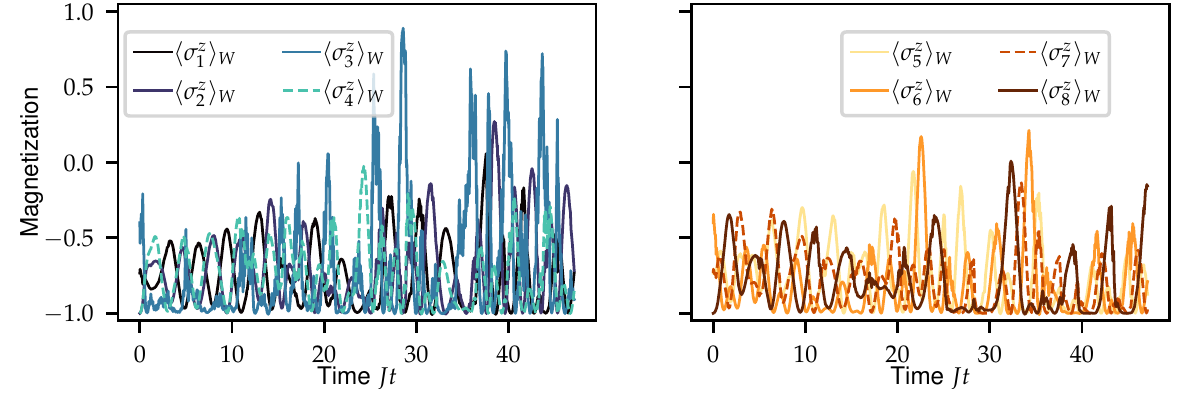}}; 
  \node (c) [label={[label distance=-0.6 cm]162: \textbf{(c)}}] at (0,-8.8) {\includegraphics{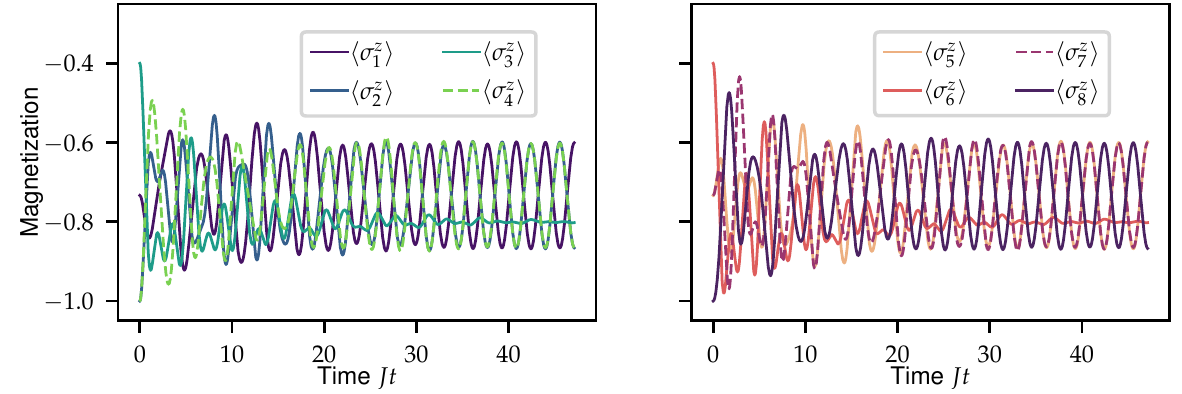}}; 
  \draw (a) node[xshift=-1.4cm,yshift=2.27cm] {\textsf{Trajectory level (first decoherence-free subspace)}}; 
  \draw (b) node[xshift=-1.9cm,yshift=2.23cm] {\textsf{Trajectory level (orthogonal complement)}};
  \draw (c) node[xshift=-3.7cm,yshift=2.23cm] {\textsf{Ensemble level}};
  \end{tikzpicture}
  \caption{Evolution of  all the magnetizations $\langle\sigma^z_i \rangle_W$ of the $XY$ spin chain, \cref{chain}, with $N=8$ spins subjected to homodyne measurement of the third $z$-polarization with measurement operator $L = \sqrt{\Gamma} \sigma^z_3$. 
  The left columns  respectively show the magnetizations of the first half of the chain and the right columns those  of second half of the chain  (in analogy to Figs.~1a-c of the main text but for all the magnetizations). In the first decoherence-free subspace, the first qubit oscillates in anti-phase with the second, the fourth and the eighth qubits, whereas it oscillates in phase with the fifth and the seventh qubits (qubits one and six have zero amplitude); {(a) and (b)} display the trajectory level and (c) the ensemble level. (b) In the orthogonal complement, magnetizations oscillate randomly and no (anti)synchronization occurs.}
  \label{s1}
\end{figure}
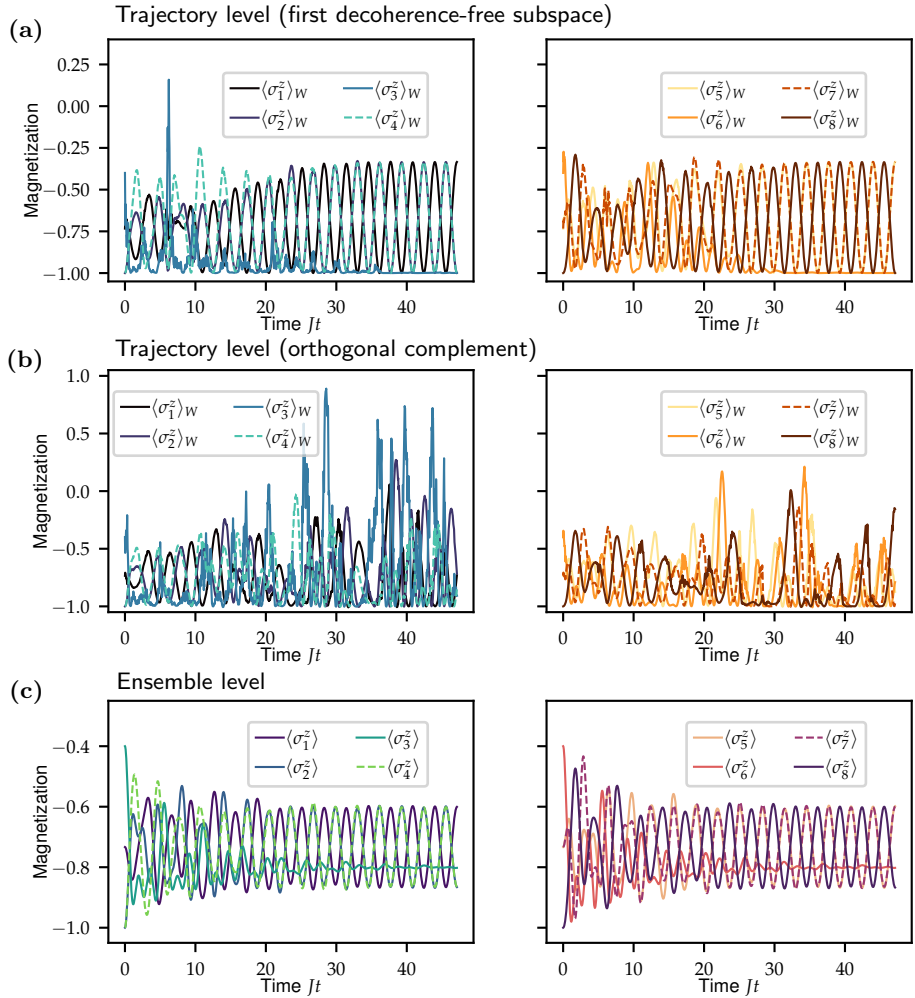
\newpage
\section{Synchronization behavior of other magnetizations}
In Fig.~1 of the main text, we only showed the (anti)synchronization properties of the two edge-spin magnetizations of the quantum chain for better visibility.
We here discuss the dynamics of all the magnetizations. We again consider the first decoherence-free subspace that supports a single eigenfrequency.
This ensures that, whenever (anti)synchronization occurs (for individual trajectories as well as for the ensemble average), there is only one eigenmode $\ket{\epsilon_\mathrm{sync}}$ in the chain corresponding to this frequency.
This eigenmode can be computed explicitly with the help of the Jordan-Wigner transformation (see Eq.~(9) in Ref.~\cite{sch22}). For the example of Fig.~1 (with $N=8$), it is given by
\begin{align}
    \ket{\epsilon_\mathrm{sync}} = \frac{3}{N+1}(1,-1,0,-1,1,0,1,-1)^\mathrm{T}.
\end{align}
The absolute values of the entries denote the amplitude of the oscillation and the sign is associated with the relative phases of the oscillating local observables $\langle \sigma^z_j\rangle$ in this eigenmode.
Consequently, the first qubit will oscillate in anti-phase with the second, the fourth and the eighth qubits, whereas it will oscillate in phase with the fifth and the seventh qubits.
Qubits one and six have zero amplitude and will therefore reach an asymptotic constant value. This behavior is illustrated in Fig.~\ref{s1}a at the trajectory level and in Fig.~\ref{s1}c at the ensemble level.

Whenever a trajectory gets trapped in the orthogonal complement, the pathwise ergodic theorem of Ref.~\cite{kum04} ensures that the trajectory will explore the whole subspace. Since this orthogonal complement is not decoherence-free, the time evolution all of the magnetizations inside the subspace remains random, and multiple eigenmodes and eigenfrequencies are excited at the same time with randomly changing weights.
For this reason, (anti) synchronization does not occur, and every spin undergoes stochastic evolution as seen in Fig.~\ref{s1}b.

\begin{figure}[t]
\centering
  \begin{tikzpicture}
    \node (a) [label={[label distance=-0.2 cm]130: \textbf{(a)}}] at (0,0) {\includegraphics{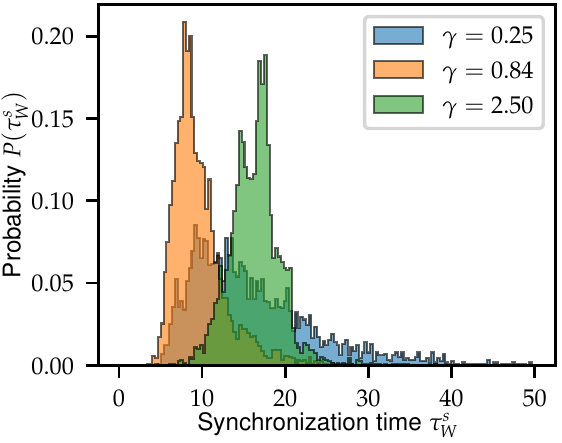}};	
    \node (b) [label={[label distance=-0.2 cm]130: \textbf{(b)}}] at (5.7,0) {\includegraphics{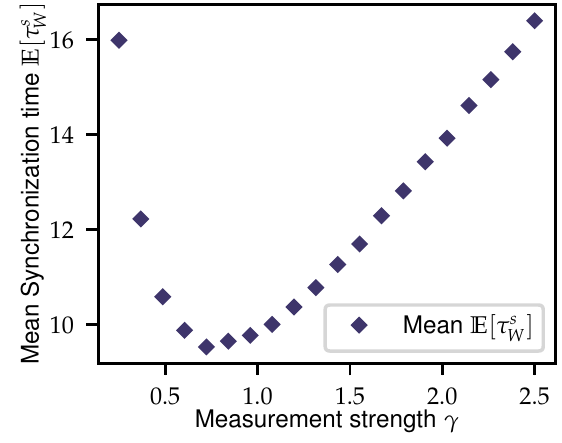}}; 
    \node (c) [label={[label distance=-0.2 cm]130: \textbf{(c)}}] at (11.4,0) {\includegraphics{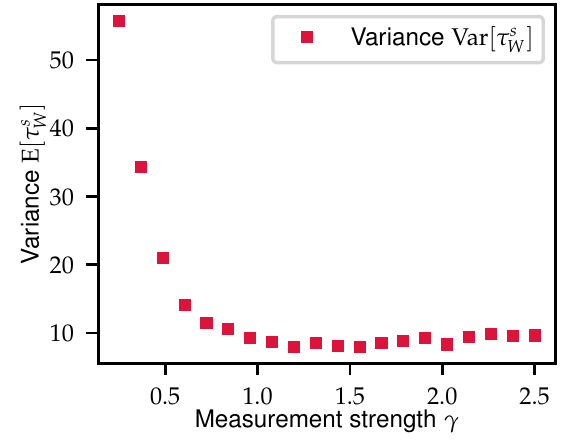}};	
  \end{tikzpicture}
  \caption{Synchronization time $\tau^s_W$ as a function of the reduced measurement strength $\gamma$ for a  quantum chain of $N = 5$ qubits with Hamiltonian \eqref{seq:chain}. This model features a decoherence-free subspace with the same eigenmode and eigenfrequency as the spin chain considered in the main text with $N = 8$. The initial state is an equal superposition of a decoherence-free subspace and the orthogonal complement $\ket{\Psi(0)} = (\ket{q_2}+\ket{p})/\sqrt{2}$. 
    (a) Histogram of the synchronization time $\tau^s_W$ for different values of the reduced measurement strength.
    (b) Mean synchronization time $\mathbb{E}[\tau^s_W]$ as a function of $\gamma$.
    For strong measurement, synchronization takes longer on average due to the quantum Zeno effect.
    (c) Variance of the synchronization time as a function of $\gamma$.
    The variance is large for weak measurements and decreases with increasing $\gamma$ until reaching a finite asymptotic value in the quantum Zeno regime.
    For each value of $\gamma$, $7000$ trajectories were simulated.}
  \label{fig:sync-time}  
\end{figure}

\section{Synchronization time scale}
In this section, we investigate the time scale of measurement-induced synchronization, which mainly depends on the reduced measurement strength $\gamma = \Gamma/J$ and on the system size $N$. 
Since individual trajectories are described by a stochastic process, the synchronization time $\tau^s_W$ will be a random quantity (indicated by the index $W$).
The probability for the system to be found in the decoherence-free subspace, Eq.~\eqref{seq:dq}, follows a free Brownian motion with state-dependent diffusion, \cref{seq:sse}, that terminates at the boundaries (i.e. whenever $|q|^2 = \{0,1\}$).
For a single trajectory, the synchronization time is therefore given by the first hitting time $\tau^s_W$ corresponding to $|q(\tau^s_W)|^2 = 1$ \cite{kam92}. 
We numerically evaluate the distribution of the synchronization time $P(\tau^s_W)$ for the quantum spin chain \eqref{seq:chain} with $N = 5$ qubits that features the same decoherence-free subspace as the model investigated in Fig.~1 of the main text (more specifically, with the same eigenfrequency and eigenmode in the synchronized regime).
We fix the initial state to be in an equal superposition of a state from the second (two-dimensional) decoherence-free subspace and a  state from a (three-dimensional) orthogonal non-decoherence-free subspace in the complement
\begin{align}
  \ket{\Psi(0)} = {\frac{1}{\sqrt{2}}}\left(\ket{q_2}+\ket{p}\right),
\end{align}
so that (anti)synchronization occurs with a probability of $|q_2(0)|^2 = 1/2$.

\Cref{fig:sync-time}a shows the distribution $P(\tau^s_W)$ of the synchronization time  for three different values of the reduced measurement strength $\gamma$, whereas \cref{fig:sync-time}b and \cref{fig:sync-time}c respectively exhibit the mean synchronization time $\mathbb{E}[\tau^s_W]$ and the variance $\text{Var}[\tau^s_W]$, both   as a function of $\gamma$.
We observe that for weak measurements (small $\gamma$), the distributions $P(\tau^s_W)$ are broad and asymmetric with longer tails towards higher values of $\tau^s_W$ (\cref{fig:sync-time}a); they become increasingly symmetric and narrow for increasing measurement strengths before saturating to a finite width (\cref{fig:sync-time}c). 
On the other hand, the mean synchronization time $\mathbb{E}[\tau^s_W]$ experiences a crossover as the strength $\gamma$ is varied  (\cref{fig:sync-time}b): it diverges in the limit of no measurements ($\gamma \to 0$), indicating that the weaker the measurement, the longer it takes for the system  to synchronize on average.
The mean synchronization time then decreases for stronger measurements until a critical value is reached, after which $\mathbb{E}[\tau^s_W]$ increases again due to the quantum Zeno effect \cite{sud77}.
We further note that the behavior for the average synchronization time is in agreement with the analytical results obtained in Ref.~\cite{sch22}  for the average evolution, where the synchronization time was extracted from the spectrum of the Liouvillian (the synchronization time was there found to display an algebraic dependence on the chain size $N$).

\section{Average fidelity}
Finally, we detail  the computation of the average fidelity between the time-averaged state $\overline{\rho}_W $ and the ensemble-averaged state $\rho^\m{s}$, Eq.~(6) of the main text. We further characterize the breaking of ergodicity by giving bounds on the mean fidelity and its variance.
Given a Lindblad equation with a set of operators ($H,L$) and an initial state $\rho(0)$, the Hilbert space can be uniquely decomposed into a series of (minimal) orthogonal subspaces \cite{bau08,bau12}.
Accordingly, the density matrix $\rho(t)$, the Lindblad jump operator $L$ and the Hamiltonian $H$ admit a simultaneous block decomposition \cite{bau08,bau12}.
Under these conditions, each minimal orthogonal subspace supports a unique asymptotic state $\rho^\m{s}_j$, and the asymptotic state of the Lindblad equation $\rho ^\m{s}$ can be written as $\rho ^\m{s} = \sum_j w_j \rho^\m{s}_j$.
We assume that the blocks of $L$ are independent such that \cref{seq:dq} has only the two fixed points $\{0,1\}$. 
Then, by the pathwise ergodic theorem \cite{kum04}, the time average of every trajectory $\rho_W$ will converge to one of the asymptotic states $\rho^\m{s}_j$ of the Lindblad equation $\overline{\rho}_{W,j} = \rho^\m{s}_j$, and the mean fidelity becomes 
\begin{equation}
  \mathbb{E}[F\left(\overline{\rho}_{W},\rho^\m{s}\right)]
    = \sum_k w_k F\bigg(\overline{\rho}_{W,k},\sum_j w_j \rho^\m{s}_j\bigg)
    = \sum_k w_k^2.
    \label{seq:ipr}
\end{equation}
\Cref{seq:ipr} is a participation ratio of the initial state with respect to the state space structure
\begin{align}
  \mathbb{E}[F\left(\overline{\rho}_{W},\rho^\m{s}\right)]
  = R 
  = \sum_k \Tr[\Pi_k \rho(0)]^2, 
\end{align}
where $\Pi_k$ is a projector on the $k${th} subspace.
The average fidelity is lower bounded by the uniform distribution
\begin{align}
  \mathbb{E}[F\left(\overline{\rho}_{W},\rho^\m{s}\right)] \ge \frac{1}{N},
\end{align}
where $N$ denotes the number of orthogonal subspaces. This is, in a sense, the maximal possible ergodicity breaking for a given quantum system with Hamiltonian $H$ that is continuously monitored via the measurement operator $L$. This is achieved for instance in Fig.~1 of the main text (lower panel).
The average fidelity only depends on the number of orthogonal subspaces and the initial distribution of the density matrix over them.
It is, however, independent from their internal structure and dimension.
The ergodicity breaking is thus determined by the delocalization of the initial condition over the orthogonal subspaces as measured by the corresponding participation ratio $R$.
The more the Hilbert space is fractionalized, the more localized individual realizations can become with respect to the behavior of the average.
In the case of equal weights in each subspace $w_j = 1/N$, the variance
\begin{align}
  \m{Var}[F]
  = \sum_{k=1}^N w_k^3 - \left[\sum_{j=1}^N w_j^2\right]^2.
\end{align}
vanishes.
Finally, we can employ Popoviciu's inequality \cite{mit93} to upper bound the variance
\begin{align}
  \m{Var}[F] \le \frac{[\m{min}(F)-\m{max}(F)]^2}{4} = \frac{(1/N-1)^2}{4}.
\end{align}


\begin{thebibliography}{99}
   \bibitem{ble88} I. I. Blekman, \textit{Synchronization in Science and Technology}, (ASME Press, New York, 1988).
   \bibitem{boc02} S. Boccaletti, J. Kurths, G. Osipov, D. L. Valladares, C. S. Zhou, The synchronization of chaotic systems, Phys. Rep. \textbf{366} (2002).
   \bibitem{pik03} A. Pikovsky, M. Rosenblum, and J. Kurths, \textit{Synchronization: A Universal Concept in Nonlinear Sciences} (Cambridge University Press, Cambridge, 2003).
   \bibitem{ace05} J. A. Acebr\'on, L. L. Bonilla, C. J. P\'erez Vicente, F. Ritort, and R. Spigler, The Kuramoto model: A simple paradigm for synchronization phenomena, Rev. Mod. Phys. \textbf{77}, 137 (2005).
   \bibitem{mos02} E. Mosekilde, Y. Maistrenko, and D. Postnov, \textit{Chaotic Synchronization}, (World Scientific, Singapore, 2002). 
   \bibitem{ani07} V. S. Anishchenko, V. Astakhov, A. Neiman, T. Vadivasova, and L. Schimansky-Geier, \textit{
   Nonlinear Dynamics of Chaotic and Stochastic Systems}, (Springer, Berlin, 2007). 
   \bibitem{bal09} A. Balanov, N. Janson, D. Postnov, and O. Sosnovtseva, \textit{Synchronization. From Simple to Complex}, (Springer, Berlin, 2009).
   \bibitem{boc18} S. Boccaletti, A. N. Pisarchik, C. I. del Genio, and A. Amann, \textit{Synchronization. From Coupled Systems to Complex Networks}, (Cambridge University Press, Cambridge, 2018).
   
   \bibitem{goy06} I. Goychuk, J. Casado-Pascual, M. Morillo, J. Lehmann, and P. H\"anggi,
    Quantum Stochastic Synchronization, Phys. Rev. Lett. \textbf{97}, 210601 (2006).
   \bibitem{zhi08} O. V. Zhirov and D. L. Shepelyansky, Synchronization and Bistability of a Qubit Coupled to a Driven Dissipative Oscillator, Phys. Rev. Lett. \textbf{100}, 014101 (2008).
   \bibitem{hei11} G. Heinrich, M. Ludwig, J. Qian, B. Kubala, and F. Marquardt, Collective Dynamics in Optomechanical Arrays, Phys. Rev. Lett. \textbf{107}, 043603 (2011).
   
   \bibitem{lud13} M. Ludwig and F. Marquardt, Quantum Many-Body Dynamics in Optomechanical Arrays, Phys. Rev. Lett. \textbf{111}, 073603 (2013).
   \bibitem{mar13} A. Mari, A. Farace, N. Didier, V. Giovannetti, and R. Fazio, Measures of Quantum Synchronization in Continuous Variable Systems, Phys. Rev. Lett. \textbf{111}, 103605 (2013).
   \bibitem{lee13} T. E. Lee and H. R. Sadeghpour, Quantum Synchronization of Quantum van der Pol Oscillators with Trapped Ions, Phys. Rev. Lett. \textbf{111}, 234101 (2013).
   \bibitem{wal14} S. Walter, A. Nunnenkamp, and C. Bruder, Quantum Synchronization of a Driven Self-Sustained Oscillator,
   Phys. Rev. Lett. \textbf{112}, 094102 (2014).
   \bibitem{lor16} N. L\"orch, E. Amitai, A. Nunnenkamp, and C. Bruder, Genuine Quantum Signatures in Synchronization of Anharmonic Self-Oscillators, Phys. Rev. Lett. \textbf{117}, 073601 (2016).
   \bibitem{lor17} N. L\"orch, S. E. Nigg, A. Nunnenkamp, R. P. Tiwari, and C. Bruder, Quantum Synchronization Blockade: Energy Quantization Hinders Synchronization of Identical Oscillators, Phys. Rev. Lett. \textbf{118}, 243602 (2017).
   \bibitem{rou18} A. Roulet and C. Bruder, Quantum Synchronization and Entanglement Generation, Phys. Rev. Lett. \textbf{121}, 063601 (2018).
   \bibitem{rou18a} A. Roulet and C. Bruder, Synchronizing the Smallest Possible System, Phys. Rev. Lett. \textbf{121}, 053601 (2018).
   \bibitem{son18} S. Sonar, M. Hajdusek, M. Mukherjee, R. Fazio, V. Vedral, S. Vinjanampathy, and L.-C. Kwek, Squeezing Enhances Quantum Synchronization, Phys. Rev. Lett. \textbf{120}, 163601 (2018).
   \bibitem{cab19} A. Cabot, G. L. Giorgi, F. Galve, and R. Zambrini, Quantum Synchronization in Dimer Atomic Lattices, Phys. Rev. Lett. \textbf{123}, 023604 (2019).
   \bibitem{las20} A. W. Laskar, P. Adhikary, S. Mondal, P. Katiyar, S. Vinjanampathy, and S. Ghosh,
   Observation of Quantum Phase Synchronization in Spin-1 Atoms, Phys. Rev. Lett. \textbf{125}, 013601 (2020).
   \bibitem{sch22} F. Schmolke and E. Lutz, Noise-Induced Quantum Synchronization, Phys. Rev. Lett. \textbf{129}, 250601 (2022).
   
   \bibitem{wis09} H. M. Wiseman and G. J. Milburn, \textit{Quantum Measurement and Control}, (Cambridge University Press, Cambridge, 2009). 
   \bibitem{bar09}A. Barchielli and M. Gregoratti, \textit{Quantum Trajectories and Measurements in Continuous Time} (Springer, Berlin, 2009).
   \bibitem{jac14} K. Jacobs, \textit{Quantum Measurement Theory}, (Cambridge University Press, Cambridge, 2014).
   \bibitem{sud77} E. C. G. Sudarshan and B. Misra, The Zeno's paradox in quantum theory, J. Math. Phys. \textbf{18} 756 (1977). 
   
   
   \bibitem{choi20} S. Choi, Y. Bao, X.-L. Qi, and E. Altman, Quantum error correction in scrambling dynamics and measurement-
   induced phase transition, Phys. Rev. Lett. 125, 030505 (2020).
   \bibitem{szy20} M. Szyniszewski, A. Romito, and H. Schomerus, Universality of Entanglement Transitions from Stroboscopic to Continuous Measurements, Phys. Rev. Lett. \textbf{125}, 210602 (2020).
   \bibitem{alb21} O. Alberton, M. Buchhold, and S. Diehl, Entanglement Transition in a Monitored Free-Fermion Chain: From Extended Criticality to Area Law, Phys. Rev. Lett. \textbf{126}, 170602 (2021).
   \bibitem{gop21} S. Gopalakrishnan and M. J. Gullans, Entanglement and Purification Transitions in Non-Hermitian Quantum Mechanics,
   Phys. Rev. Lett. \textbf{126}, 170503 (2021).
   \bibitem{jia21} S.-K. Jian, C. Liu, X. Chen, B. Swingle, and P. Zhang, Measurement-Induced Phase Transition in the Monitored Sachdev-Ye-Kitaev Model, Phys. Rev. Lett. \textbf{127}, 140601 (2021).
   \bibitem{vov22} T. Vovk and H. Pichler, Entanglement-Optimal Trajectories of Many-Body Quantum Markov Processes, Phys. Rev. Lett. \textbf{128}, 243601 (2022).
   \bibitem{sie22} P. Sierant and X. Turkeshi, Universal Behavior beyond Multifractality of Wave Functions at Measurement-Induced Phase Transitions, Phys. Rev. Lett. \textbf{128}, 130605 (2022).
   \bibitem{zab22} A. Zabalo, M. J. Gullans, J. H. Wilson, R. Vasseur, A. W. W. Ludwig, S. Gopalakrishnan, David A. Huse, and J. H. Pixley, Operator Scaling Dimensions and Multifractality at Measurement-Induced Transitions, Phys. Rev. Lett. \textbf{128}, 050602 (2022).
   \bibitem{mul22} T. M\"uller, S. Diehl, and M. Buchhold, Measurement-Induced Dark State Phase Transitions in Long-Ranged Fermion Systems, Phys. Rev. Lett. \textbf{128}, 010605 (2022).
   \bibitem{blo22} M. Block, Y. Bao, S. Choi, E. Altman, and N. Y. Yao, Measurement-Induced Transition in Long-Range Interacting Quantum Circuits, Phys. Rev. Lett. \textbf{128}, 010604 (2022)
   \bibitem{min22} T. Minato, K. Sugimoto, T. Kuwahara, and K. Saito, Fate of Measurement-Induced Phase Transition in Long-Range Interactions, Phys. Rev. Lett. \textbf{128}, 010603 (2022).
   
   
   \bibitem{geb20} V. Gebhart, K. Snizhko, T. Wellens, A. Buchleitner, A. Romito, and Y. Gefen, Topological transition in measurement-induced geometric phases, PNAS \textbf{117}, 5706 (2020).
   \bibitem{sni21} K. Snizhko, P. Kumar, N. Rao, and Y. Gefen, Weak-Measurement-Induced Asymmetric Dephasing: Manifestation of Intrinsic Measurement Chirality, Phys. Rev. Lett. \textbf{127}, 170401 (2021).
   \bibitem{wan22} Y. Wang, K. Snizhko, A. Romito, Y. Gefen, and K. Murch, Observing a topological transition in weak-measurement-induced geometric phases, Phys. Rev. Research \textbf{4}, 023179 (2022).
   
   \bibitem{dua97} L.-M. Duan and G.-C. Guo, Preserving Coherence in Quantum Computation by Pairing Quantum Bits, Phys. Rev. Lett. \textbf{79}, 1953 (1997).
   \bibitem{zan97} P. Zanardi and M. Rasetti, Noiseless Quantum Codes, Phys. Rev. Lett. \textbf{79}, 3306 (1997).
   \bibitem{lid98} D. A. Lidar, I. L. Chuang, and K. B. Whaley, Decoherence-Free Subspaces for Quantum Computation,
   Phys. Rev. Lett. \textbf{81}, 2594 (1998).
   \bibitem{lid03} D. A. Lidar and K. B. Whaley, Decoherence-free subspaces and subsystems, Lecture Notes in Physics \textbf{622}, 83 (2003).
   \bibitem{blu08} R. Blume-Kohout, H. K. Ng, D. Poulin, and L. Viola, Characterizing the Structure of Preserved Information in Quantum Processes, Phys. Rev. Lett. \textbf{100}, 030501 (2008).
   
   \bibitem{joz94} R. Jozsa, Fidelity for Mixed Quantum States, J. Mod. Opt. \textbf{41}, 2315 (1994). 
   
   \bibitem{zie09} R. E. Ziemer and W. H. Tranter, \textit{Principles of Communications}, (Wiley, New York, 2009).
   \bibitem{nie00} M. A. Nielsen and I. L. Chuang, \textit{Quantum Computation and Quantum Information}, (Cambridge
   University Press, Cambridge, 2000).
   \bibitem{dut15} A. Dutta, G. Aeppli, B. K. Chakrabarti, U. Divakaran, T. F. Rosenbaum and D. Sen, \textit{Quantum Phase Transitions in Transverse Field Spin Models}, (Cambridge University Press, Cambridge, 2015).
   
   \bibitem{wei16} T. Weiss, A. Kronwald, and F. Marquardt, Noise-induced transitions in optomechanical synchronization, New J. Phys. \textbf{18}, 013043 (2016)
   \bibitem{es20} N. Es'haqi-Sani, G. Manzano, R. Zambrini, and R. Fazio, Synchronization along quantum trajectories, Phys. Rev. Res. \textbf{2}, 023101 (2020).
   \bibitem{gio12} G. L. Giorgi, F. Galve, G. Manzano, P. Colet, and R. Zambrini, Quantum correlations and mutual synchronization, Phys. Rev. A \textbf{85}, 052101 (2012).
   \bibitem{gio13} G. L. Giorgi, F. Plastina, G. Francica, and R. Zambrini, Spontaneous synchronization and quantum correlation dynamics of open spin systems, Phys. Rev. A \textbf{88}, 042115 (2013).
   \bibitem{kar19} G. Karpat, I. Yalcinkaya, and B. Cakmak, Quantum synchronization in a collision model, Phys. Rev. A \textbf{100}, 012133 (2019).
   \bibitem{gio19} G. L. Giorgi, A. Cabot, and R. Zambrini, Transient synchronization in open quantum systems, Springer Proceedings in Physics \textbf{237}, 73 (2019).
   \bibitem{kar20} G. Karpat, I. Yalcinkaya, and B. Cakmak, Quantum synchronization of few-body systems under collective dissipation,
   Phys. Rev. A \textbf{101}, 042121 (2020).
   \bibitem{tyn20} J. Tindall, C. S. Munoz, B. Buca, and D. Jaksch, Quantum synchronisation enabled by dynamical symmetries and dissipation, New J. Phys. \textbf{22}, 013026 (2020).
   \bibitem{buc22} B. Buca, C. Booker, and D. Jaksch, Algebraic theory of quantum synchronization and limit cycles under dissipation, SciPost Phys. \textbf{12}, 097 (2022).
   
   \bibitem{com} This condition is not necessary, however, as stable quantum synchronization may also occur when the decoherence-free subspaces contain multiple frequencies, as seen in the multiplexing example.
   
   \bibitem{sup} See Supplemental Material for  details about the dynamics of the quantum synchronization process, including the determination of the decoherence-free subspaces and the characterization of ergodicity breaking, which includes Refs.~\cite{str17,sin15,gya20,mit93}. 
   
   \bibitem{str17} A. Streltsov, G. Adesso, and M. B. Plenio, Quantum coherence as a resource,  Rev.  Mod. Phys.  \textbf{89}, 041003 (2017).
   \bibitem{sin15} U. Singh, M. Nath Bera, H. S. Dhar, and A. K. Pati, Maximally coherent mixed states: Complementarity between maximal coherence and mixedness, Phys. Rev. A \textbf{91}, 052115 (2015).
   \bibitem{gya20} J. A. Gyamfi, Fundamentals of quantum mechanics in Liouville space, Eur. J. Phys. \textbf{41}, 063002 (2020).
   \bibitem{mit93} D. S. Mitrinovic, J. E. Pecaric, and A. M. Fink, \textit{Classical and New Inequalities in
   Analysis}, (Springer, Berlin, 1993).
   
    
   \bibitem{kam92} N. G. Van Kampen, \textit{Stochastic processes in Physics and Chemistry} (Elsevier, Amsterdam, 1992).
   
   
   
   \bibitem{san19} C. S\'anchez Mun\~oz, B. Buca, J. Tindall, A. Gonz\'alez- Tudela, D. Jaksch, and D. Porras, Symmetries and conservation laws in quantum trajectories: Dissipative freezing, Phys. Rev. A \textbf{100}, 042113 (2019).
   \bibitem{hal22} C.-M. Halati, A. Sheikhan, and C. Kollath, Breaking strong symmetries in dissipative quantum systems: Bosonic atoms coupled to a cavity, Phys. Rev. Res. \textbf{4}, L012015 (2022).
   \bibitem{tin23} J. Tindall, D. Jaksch, and C. S. Mun\~oz,
   On the generality of symmetry breaking and dissipative freezing in quantum trajectories, SciPost Phys. Core \textbf{6}, 004 (2023).
   
   
   
   
   
   
   \bibitem{guo21} X. Y. Guo, Z. Y. Ge, H. Li \textit{et al.}, Observation of Bloch oscillations and Wannier-Stark localization on a superconducting quantum processor, npj Quantum Inf. \textbf{7}, 51 (2021). 
   \bibitem{mi22} X. Mi, M. Sonner, M. Y. Niu, K. W. Lee \textit{et al.}, Noise-resilient edge modes on a chain of superconducting qubits, Science \textbf{378}, 785 (2022).
   
   \bibitem{mur13} K. W. Murch, S. J. Weber, K. M. Beck, E. Ginossar, and I. Siddiqi, Reduction of the radiative decay of atomic coherence in squeezed vacuum, Nature \textbf{499}, 62 (2013).
   \bibitem{web14} S. J. Weber, A. Chantasri, J. Dressel, A. N. Jordan, K. W. Murch, and I. Siddiqi, Mapping the optimal route between two quantum states, Nature \textbf{511}, 570 (2014).
   
   
   \bibitem{pal82} R. G. Palmer, Broken ergodicity, Adv. Phys. \textbf{31}, 669 (1982).
   \bibitem{bau08} B. Baumgartner and H. Narnhofer, Analysis of quantum semigroups with GKS Lindblad generators: II. General, J. Phys. A \textbf{41}, 395303 (2008).
   \bibitem{bau12} B. Baumgartner and H. Narnhofer, The structures of state space concerning quantum dynamical semigroups, Rev. Math. Phys. \textbf{24}, 1250001 (2012).
   \bibitem{kum04} B. K\"ummerer and H. Maassen, A pathwise ergodic theorem for quantum trajectories, J. Phys. A \textbf{37}, 11889 (2004).
   
   \bibitem{car16} R. Carbone and Y. Pautrat, Irreducible Decompositions and Stationary States of Quantum Channels, Rep. Math. Phys., \textbf{77}, 3 (2016).
   
   \bibitem{com1} The theorems of Refs.~\cite{bau08,bau12} are valid for finite-dimensional systems, like spin systems. However, they have recently been extended to infinite-dimensional systems in Ref.~\cite{car16}, indicating that measurement-induced synchronization may also emerge in the latter.
   
   \bibitem{kra93} B. Kramer and A. MacKinnon, Localization: theory and experiment, Rep. Prog. Phys. \textbf{56}, 1469 (1993).
   \bibitem{eve08} F. Evers and A. D. Mirlin, Anderson transitions, Rev. Mod. Phys. \textbf{80}, 1355 (2008).
   
   \bibitem{arg05} A. Argyris, D. Syvridis, L. Larger, V. Annovazzi-Lodi, P. Colet, I. Fischer, J. Garca-Ojalvo, C.R. Mirasso, L. Pesquera, and K. A. Shore, Chaos-based communications at high bit rates using commercial fibre-optic links. Nature \textbf{438}, 343346 (2005).
   
   \bibitem{cho17} H.-H. Choi and J.-R. Lee, Principles, Applications, and Challenges of Synchronization in Nature for Future Mobile Communication Systems, Mobile Inf. Syst. \textbf{2017}, 8932631
   (2017).
   
   
   
   
   \end{thebibliography}

\newpage
\begin{thebibliography}{99}
\bibitem{wis09} H. M. Wiseman and G. J. Milburn, \textit{Quantum Measurement and Control}, (Cambridge University Press, Cambridge, 2009). 
\bibitem{bar09}A. Barchielli and M. Gregoratti, \textit{Quantum Trajectories and Measurements in Continuous Time} (Springer, Berlin, 2009).
\bibitem{jac14} K. Jacobs, \textit{Quantum Measurement Theory}, (Cambridge University Press, Cambridge, 2014).


\bibitem{str17} A. Streltsov, G. Adesso, and M. B. Plenio, Quantum coherence as a resource,  Rev.  Mod. Phys.  \textbf{89}, 041003 (2017).
\bibitem{sin15} U. Singh, M. Nath Bera, H. S. Dhar, and A. K. Pati, Maximally coherent mixed states: Complementarity between maximal coherence and mixedness, Phys. Rev. A \textbf{91}, 052115 (2015).

\bibitem{kam92} N. G. Van Kampen, \textit{Stochastic processes in Physics and Chemistry} (Elsevier, Amsterdam, 1992).
\bibitem{gya20} J. A. Gyamfi, Fundamentals of quantum mechanics in Liouville space, Eur. J. Phys. \textbf{41}, 063002 (2020).

\bibitem{sch22} F. Schmolke and E. Lutz, Noise-Induced Quantum Synchronization, Phys. Rev. Lett. \textbf{129}, 250601 (2022).

\bibitem{sud77} E. C. G. Sudarshan and B. Misra,  The Zeno's paradox in quantum theory, J.  Math. Phys. \textbf{18},  756 (1977). 

\bibitem{bau08} B. Baumgartner and H. Narnhofer, Analysis of quantum semigroups with GKS Lindblad generators: II. General, J.  Phys. A \textbf{41}, 395303 (2008).
\bibitem{bau12} B. Baumgartner and H. Narnhofer, The structures of state space concerning quantum dynamical semigroups, Rev.  Math. Phys. \textbf{24}, 1250001 (2012).
\bibitem{kum04} B. K\"ummerer and H. Maassen, A pathwise ergodic theorem for quantum trajectories, J. Phys. A \textbf{37}, 11889 (2004).
\bibitem{mit93} D. S. Mitrinovic, J. E. Pecaric, and A. M. Fink, \textit{Classical and New Inequalities in
Analysis}, (Springer, Berlin, 1993).
\end{thebibliography}
\end{document}